# Integrated Modeling and Forecasting of Electric Vehicles Charging Profiles Based on Real Data

Octavio Ramos-Leaños *
Hussein Suprême *
Mouhamadou Makthar Dione *
* Contrôle et gestion des réseaux
* Institut de recherche d'Hydro-Québec
Varennes, Québec, Canada
ramos.octavio@hydroquebec.com

Daniel Chabot ¥
Vincent Beaulieu ¥

¥ Hydro-Québec
¥ Montréal, Québec, Canada
supreme.hussein2@hydroquebec.com

*Abstract—* In the context of energy transition and decarbonization of the economy, several governments will ban the sale of new combustion vehicles by 2050. Thus, growing penetration of electric vehicles (EVs) in distribution networks (DN) is predicted. Impact analyzes must be performed to determine if mitigation means are needed to accommodate a large quantity of EVs in the DN. Furthermore, the habits of the local population resulting in different EVs charging patterns needs to be realistically considered. This article proposed an individual residential EVs multi-charging algorithm based on the observed charging behavior of 500 measured residential EVs located in a large North American utility. Probability functions are derived from the analysis of these charging patterns. These can be used to model daily charging profiles of individual EV to assess, in a quasi-static time-series perspective, their impact either on a single costumer or a whole DN. An impact evaluation study is also presented.

## I. INTRODUCTION

With electric vehicles (EVs) being exponentially adopted in some regions of the world [1], utilities must develop new methods to assess the impact of the high penetration of EVs in distribution networks (DN) [2]. Several publications [3]-[4]-[5] have already demonstrated the need to conduct impact studies of EVs charging on low-voltage networks in terms of steady-state voltage limit violations, system frequency, grid load capacity, power quality, feeder and transformer thermal loading, and asset degradation. Two major effects of the large integration of EVs on the grid are expected: increasing the overall power load required and increasing the load on local DNs. Power grids may trigger extreme surges in demand during rush hours, thereby harming the stability and security of existing power grids. Therefore, for studies focused on granular data to be meaningful and to provide real insights into network planners, EVs charging profiles must accurately represent local electrical load curve patterns (e.g., knowing the energy consumed by a single customer).

The charging patterns of EVs and their corresponding electrical load patterns are assessed using either a deterministic method or a probabilistic, stochastic approach. Deterministic methods can easily be derived and implemented. However, they did not consider the uncertainties related to the behavior of EV users, which affects the load pattern. Probabilistic models of EVs charging patterns consider various factors, including battery capacity, state of charge (SOC), driving habit/need, plug-in time, mileage, recharging frequency per day, charging power rate, and dynamic EV charging price under controlled and uncontrolled charging schemes. Owing to the strict confidentiality and privacy protection rules, the data required for stochastic approaches are difficult to access. Most studies in the literature used data obtained from government studies of nationwide average arrival times, traffic conditions, and distances traveled [6]-[7], surveys of owners of electric vehicles [8]-[9], and agent activity-based models [10]-[11]. On-spot measurements are a reliable source of user behavior regarding the charging of electric vehicles. However, several works used data from public charging stations and parking lots [12]-[13]. The main weakness of studies based on these data is that they do not directly represent residential charging behaviors and the number of vehicles used is very low (≤100). Charging profiles are distorted by collecting data over a short period of time using a single-vehicle model. Seasonality characteristics were also lost.

In the literature, most studies have focused on the impact of EV integration on low-voltage distribution systems. Load curves are projected as an effective way of realizing the impact and suggesting performance improvement and optimization strategies A few studies have used a stochastic approach based on real data to model the charging profiles of individual EVs. In [14], a Monte Carlo simulation (MCS)-based stochastic model was applied to develop a probabilistic model of EVs charging patterns using transportation survey data obtained from the American Time Use Survey 2009 (ATUS). These data consider several factors such as vehicle class, battery capacity, state of charge (SOC), driving habits/needs (i.e., trip type and purpose, plug-in time, mileage, recharging frequency per day, charging power rate, and dynamic EV charging price) under controlled and uncontrolled charging schemes. In [15], the authors modeled EV charging profiles based on pseudo-real driving patterns and user behavior. The inputs were based on the historical driving characteristics of private conventional vehicles from Denmark and the home plug-in behavior of EVs from Japan. The first input is used to define properties, such as the daily driven distance and the expected departure and arrival times, which determine the possible home charging window. The second input was used to quantify the probability of performing a daily domestic charge. Based on empirical mobility statistics and customizable assumptions, the open-source tool emobpy [16] can be used to create profiles of EVs charging. It consists of four time series: i) vehicle mobility containing the vehicle's location and distance traveled; ii) driving electricity consumption, specifying how much electricity is taken from the battery for driving; iii) EVs grid availability, providing information on whether and with which EV power rating is connected to the electricity grid at a certain point in time; and iv) EVs grid electricity demand, specifying the actual charging electricity drawn from the grid, based on different charging strategies. Several articles have defined an optimization problem and presented smart charging planning methods with a numerically continuous or discrete charge rate for optimum EVs integration in the DN [17] [18] [19] [20]. Some papers proposed full EVs charging, whereas others proceeded to charging based on upcoming trips and EVs initial SOCs. In these studies, details on how the charging profiles of EVs are created are still missing.

In this article, a stochastic integrated modeling and forecasting approach based on real-world charging data was proposed to better represent individual residential EV charging profiles. This approach considers low-, medium-, and long-range battery EVs and avoids computation of the battery state of charge (SOC) or EV travelling distances. Probability functions for the number of daily charging events, charging event duration, and start of the charging event were used. The latter are obtained from the analysis of charging patterns of 500 residential charging stations in Hydro-Québec (HQ), a large North American utility. The charging profiles obtained by the proposed method are then used alongside Advanced Metering Infrastructure (AMI) readings to simulate the EV penetration impacts in the DN.

The main contributions of this article are as follows:
- Introducing an EV charging model that can consider multiple charging events, different EV capacities, and local EV charging behavior.
- Avoiding of the computation of SOC or EV travelling distance.
- Modeling and forecasting based on residential consumption measurements to capture realistic local behaviors of EVs users.
- Statistical analysis of the impact of database size on modeling and forecasting EVs profiles using the proposed algorithm.

The remainder of this paper is organized as follows. The modeling and forecasting of the EVs charging patterns are described in Section II. Section III describes the proposed parameterization model and the calculations required to obtain the required inputs to feed the generator algorithm. Section IV presents the implementation of the different algorithms, case studies, and discussion. This paper is concluded in Section V.

## II. PROPOSED ALGORITHM FOR EVs CHARGING PROFILES MODELING AND FORECASTING

### A. EVs charging profiles modeling

As presented in Figure 1, the generic lithium-ion battery charging pattern is the most common and used EV charging event model in the literature. Knowledge of the maximum-power $P_{EV}$ called by the EV is required. It is limited either by the onboard charge or the charging station capacity, charging start time $t_s$, constant fast charging period $t_1$ determined by the EV battery capacity and energy required for charging, and low charging period $t_2$. Moreover, the EV battery charging process is assumed to be continuous once it starts until the battery reaches full capacity.

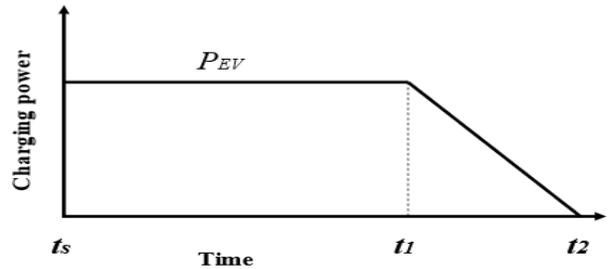

Figure 1. Generic charging profile of lithium-ion EV batteries

The previous charging pattern does not reflect the random behavior of an EV owner. Indeed, the charging process could occur at several different times in a day without reaching the maximum SOC of the battery. Furthermore, studies on the impact of EVs charging are used to mitigate the adverse effects and support the overall power system and DN. For these studies to be realistic, grid planners must consider non-worst-case scenarios of simultaneous charging. The multi-event charging model shown in Figure 2 was proposed to incorporate the variability in the profiles. Because full recharge is not mandatory, a low charging period was not considered.

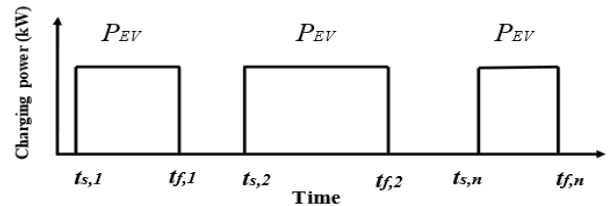

Figure 2. Illustration of an EV multi-event charging profile

The EV charging demand are determined by $P_{EV}$ the rated charging power, which is determined by the charging mode; $n$ is the number of daily charging events; $t_{s,n}$ and $t_{f,n}$ the $n$th charging event start and stop time, respectively. At time $t$, the daily charging power $C_{EV,i}$ of the $i$th EV for day $d$ can be written as (1), as the multiplication of $P_{EV,i}$ and a series of unit pulses $U_i$ for the $i$th EV.

$$C_{EV,i}(t) = P_{EV,i} \times U_i(t) \quad (1)$$

The train of unit charging events depends on the probability $\rho_{EV,i}$ of the $i$th EV charging at time $t$ and can be mathematically expressed as (2):

$$U_i(t) = \begin{cases} 0, & \rho_{EV,i}(t) = 0 \\ 1, & \rho_{EV,i}(t) \neq 0 \end{cases} \quad (2)$$

By considering the past behavior of the user, this probability can be considered as the result of three independent variables: the number of times $n$ the EV has recharged on the same day $d$, charging begins at time $t$ and the vehicle is already in charging mode at time $t$. Probabilities (3) to (5) can be associated with each of the aforementioned variables.

$$\Psi_{d,\alpha}^{rech} = \frac{\eta_{d,\alpha}}{\sum_{d \in \beta} \eta_{d,\alpha}} \quad (3)$$

where $\beta$ is the set of days in a week, $\alpha$ is the EVs type and $\eta_{d,\alpha}$ is the total number of chargings per day per type of EVs in the database.

$$\Psi_{\alpha,d,t}^{ts} = \frac{\eta_{d,\alpha,t}}{\sum_{d \in \beta} \eta_{d,\alpha,t}} \quad (4)$$

where $\eta_{d,\alpha,t}$ are the total number of charging cycles per day per type of EVs starting at time $t$ in the database.

$$\Psi_{\alpha,t}^{dur} = \frac{\eta_{\alpha,t}}{\sum_{t \in T} \eta_{\alpha,t}} \quad (5)$$

Where $\eta_{\alpha,t}$ is the total charging duration of EVs-type $\alpha$ in charging mode at $t$, and T the set of timestamps in the database.

Then, the probability $\rho_{EV,i}$ can be calculated as (6).

$$\rho_{EV,i} = \Psi_{d,\alpha}^{rech} \times \Psi_{\alpha,t}^{dur} \times \Psi_{\alpha,d,t}^{ts} \quad (6)$$

The duration of a charging event $n$ is expressed in (7), and the train of pulses $U_i(t)$ is constrained by the inequalities in (8):

$$t_{f,n} = t_{s,n} + dur_n \quad (7)$$

$$t_{f,n} > t_{s,n} > t_{f,n-1} > t_{s,n-1} > \cdots > t_{f,1} > t_{s,1} \quad (8)$$

### B. Description of the algorithm for EVs charging profiles forecasting

The model is described using the steps outlined in Algorithm 1. The inputs of the model are historical data measured by users. They are characterized by the timestamps of the measurement and the different charging powers requested by the EVs. The measured value permits the association of one of the three types of EVs with each EV. The random selection of daily charging events, start time, and charging duration introduces stochasticity in the forecasting of a user's consumption profile. The computation of the probability functions required for this algorithm is described in the following section.

**Algorithm 1.** Implementation of method for forecasting daily load profile of EV.

| Input: | Historical measurements |
|---|---|
| Output: | Single EV daily charging profile |

**Steps:**

1. For each measurement point, saving the timestamp in the following form [year-month-day hour-minute]

2. Selecting the EV's type $\alpha$ according to the charging power.

   $$\alpha = \begin{cases} small, & P_{EV} \leq 3.3\ kW \\ medium, & 3.3\ kW \leq P_{EV} \leq 7.7\ kW \\ large, & P_{EV} \geq 7.7\ kW \end{cases}$$

3. Randomly selecting number of daily charging events from the corresponding discrete probability function $\Psi_{d,\alpha}^{rech}$.

4. Randomly selecting a charging start time for each of the daily charging events, form the corresponding probability function $\Psi_{\alpha,d,t}^{ts}$.

5. Randomly selecting a charging duration for each of the daily charging events, form the corresponding probability function $\Psi_{\alpha,t}^{dur}$.

6. Computing the charging stop time $t_{f,n}$ in (7) for each charging event.

7. Addressing charging time overlaps.
   - if no overlaps: next step.
   - if overlaps: repeat 4.

8. Computing the daily charging power $C_{EV}$

### III. ANLAYSIS OF EVs CHARGING MEASURMENT DATA AND MODEL PARAMETRIZATION

By considering three types of EVs, as described in (3)–(5), nine discrete probability functions must be determined. A database was created with measurements collected from September 2018 to October 2019 from 500 charging stations for residential customers with a time step of 1 min. All study participants owned their vehicles for at least one year, and all charging points were level II chargers with a nominal voltage of 240 V and maximum current of 48 Amps. Data loggers of each station enclose:

- The charging starting time (CST) was formatted as [yyyy-mm-dd hh:mm]: the timestamp when the EV charging event started.

- The charging finishing time (CFT) was formatted as [yyyy-mm-dd hh:mm]: timestamp when the EV charging event stopped.
- Charging Consumption (CC) [kWh]: consumed energy during charging event.
- The unique identifier of each EV (EV-ID).

Then, the daily charging profiles, daily average number of charges, and average duration of charging events can be deduced.

### A. Determining the type of electric vehicles

Charging events were grouped using the EV-ID. The charging power $P_{EV,i}$ is computed using (9). Considering the EVs type defined in Algorithm 1, the fleet in the database is summarized in TABLE I.

$$P_{EV,i} = \frac{CC * 60}{CFT - CST} \quad (9)$$

TABLE I DESCRIPTION OF THE FLEET IN THE DATABASE.

| Power measured $P_{EV}$ | EVs type | Number of EVs | Ratio (%) |
|---|---|---|---|
| $P_{EV} \leq 3.3\ kW$ | small | 104 | 20.8 |
| $3.3\ kW \leq P_{EV} \leq 7.7\ kW$ | medium | 294 | 58.8 |
| $P_{EV} \geq 7.7\ kW$ | large | 102 | 20.4 |

### B. Computing probability distribution of charging duration

The charging-event duration can be computed using (7). They are grouped by EVs type to obtain the associated discrete probability density functions (DPDF) for each time step. In this study, 15 min charging cycles were adopted to represent the DPDF. Let $x$ be the charging duration of an event and $i \in \mathbb{N}$, The probability considered can be written as (10).

$$P(x_i = (i+1) \times 15) = P(i \times 15 < x_i \leq (i+1) \times 15) \quad (10)$$

In Figure 3, the 15 min PDFs of all EVs types are presented. For small-type vehicles, 15-minute charging periods were more frequent. The charging duration of more than half of the EVs was 30 min. For both types, the 95th percentile value was 75 min, and the DPDF was represented by exponential distributions. For large EVs, the charging duration was concentrated between 30 min and 90 min. Half of this fleet was charged for at least 60 minutes. The 95th percentile value was 120 minutes. The DPDF can be represented by a gamma distribution. For a larger battery capacity, users are less likely to recharge their vehicles over short periods. For all EVs types, a 45–60-minute charging duration seems to be the norm.

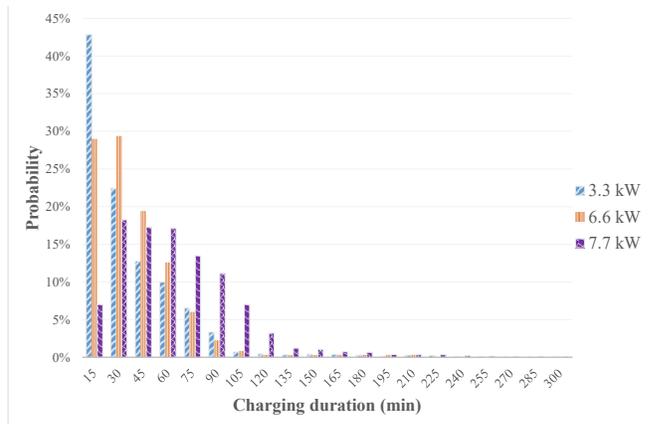

Figure 3. Discrete probability function of charging duration by EVs type

### C. Computing probability distribution of the number of daily charging events

The number of charging events per day was clustered according to the EVs type. As shown in Figure 4, for the three EVs types, there is a direct proportion of the square root of the number of daily recharges. Observing the triggering of 90% of the number of EVs in each fleet, the number of daily recharges decreased with the capacity of the batteries: five for small, three for medium, and two for large. Moreover, as shown in Figure 5, the probability of having no recharge in a day varied between 20% (small) and 30% (large).

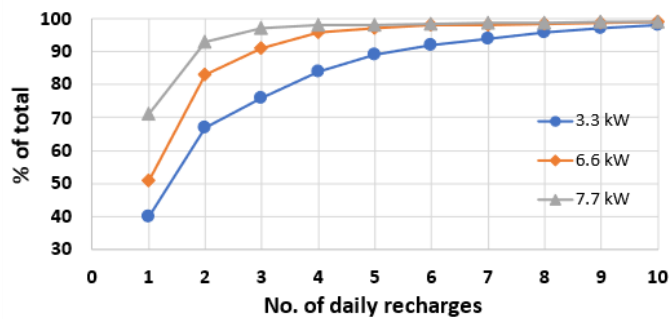

Figure 4. Cumulative percentage of daily charging events per EV type

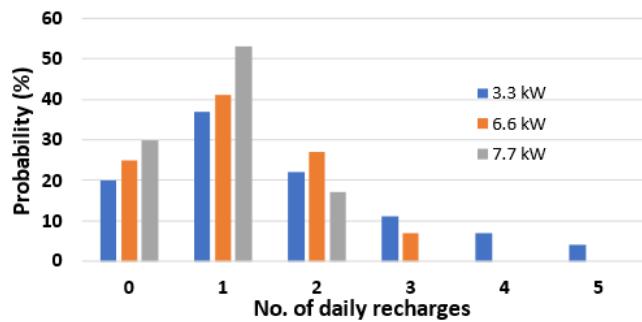

Figure 5. Discrete probability function of daily charging events by EVs type

## D. Computing probability distribution of the charging start time

All charging events measured during the year were aggregated by 24-hour period to obtain a daily hourly charging profile. With data visualization, clusters of similar EVs charging profiles across days of the week and seasons were observed. Indeed, as shown in Figure 6, user behavior follows the same trend for all Mondays in winter by maximum-normalizing the charging profile. A local peak was reached around 3:00 a.m., and the maximum peak occurred between 3:00 p.m. and 6:00 p.m. During the morning, consumption fluctuates around 0.2 p.u.

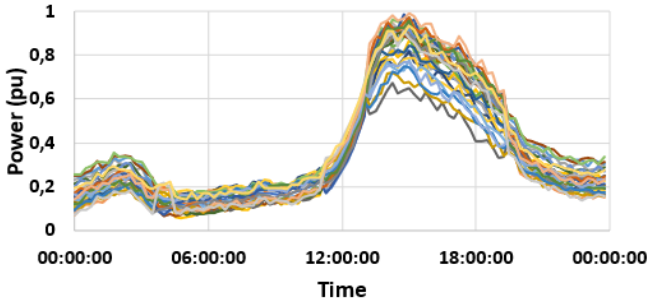

Figure 6. EVs Charging Profile – Winter Mondays

As illustrated in Figure 7.a), the load profiles thus obtained were averaged by the EVs type. For weekends, the data were merged because the profiles were nearly identical. As shown in Figure 7.b), the subsequent patterns of the charging start time are transformed into a 15 min DPDF. Medium EVs owners have a higher probability (2.6%) of starting charging at 3:00 pm on a winter Monday. Thus, the charging start probabilities are determined for all EVS types, depending on the time, day, and season.

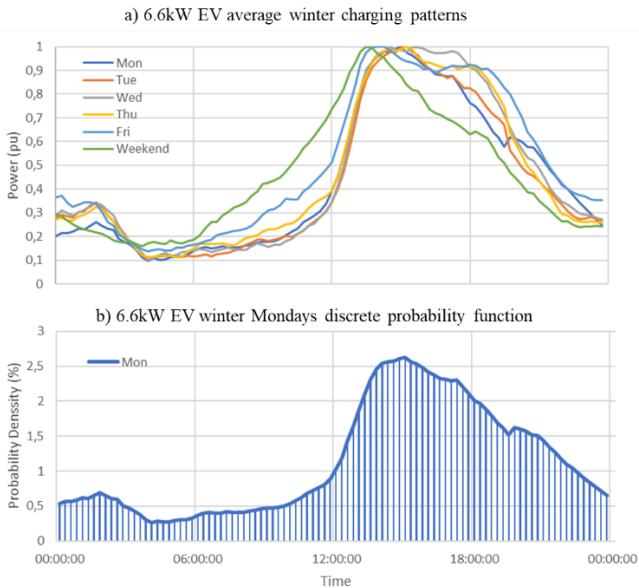

Figure 7. Average winter charging patterns for medium EVs and Mondays 15 min DPDFs illustration

## E. Importance of Database Size in Model Parameterization

As mentioned previously, studies rarely use real EV charging measurements, or a very small sample size compared to the fleet. This study used a large database of 500 EVs. To demonstrate the importance of database size, 1000 random samples of 30, 60, and 100 vehicles were chosen. As shown in Figure 8, the mean and standard deviation became significantly closer to those of the full dataset by increasing the number of EVs. Although the unavailability of measurements can distort profiles, a low-fleet sample can lead to inaccurate forecasts.

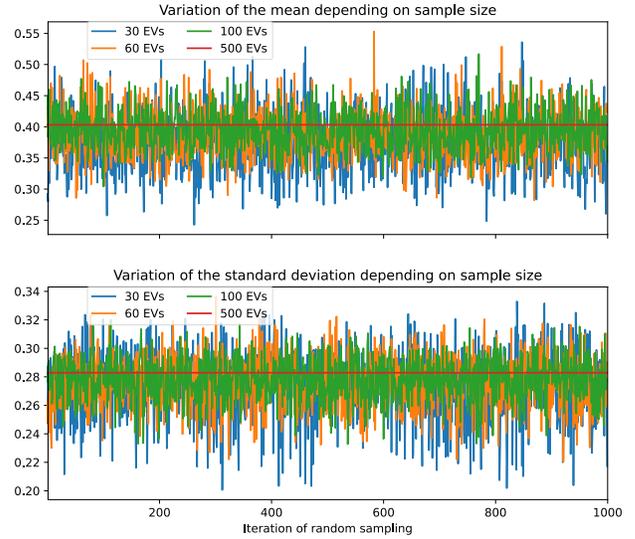

Figure 8. Variation of the mean and standard deviation for multiple random samples

Furthermore, the cosine similarity between clusters shown in Figure 9 demonstrates the importance of having a representative number of EVs to maintain cohesion in the forecasted profiles.

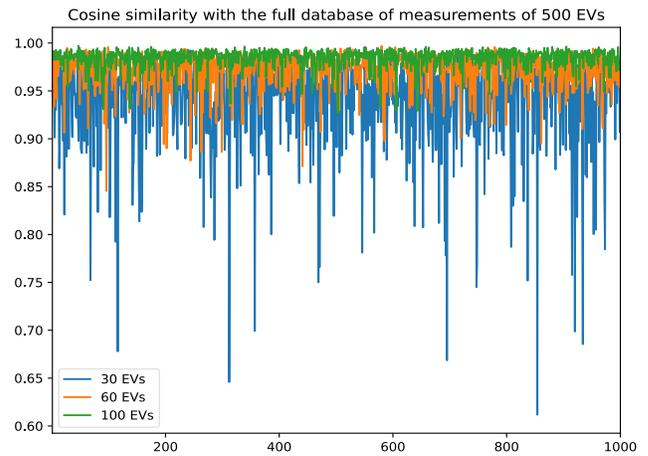

Figure 9. Similarity between profiles depending on sample size.

The deviations described above can affect the parametrization of the model. As shown in Figure 10, with a sample of 30 Vs, on a winter Monday, a pronounced valley in vehicle charging was observed between 12:00 and 15:00. In addition, the slopes around the peaks are more severe. For 60 EVs, two local peaks and a valley were observed between 05:00 and 09:00. The signal was noisy with several jumps. From the 100 EV, the charging profile is more representative of the available fleet measurements. However, the final recharge tip was not smoothed.

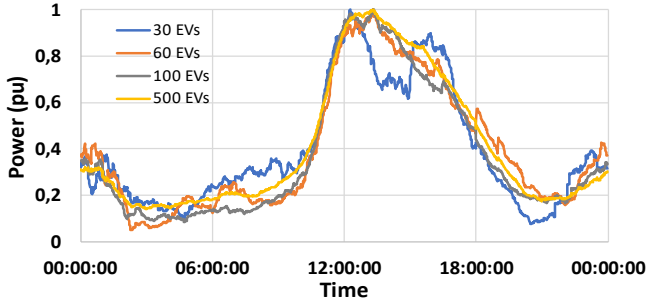

Figure 10. inter Mondays profile created from 30, 60, 100 and 500 EVs.

IV. ALGORITHM IMPLEMENTATION AND APPLICATIONS

*A. Application: creation of forecast profiles with the algorithm*

The proposed EV-charging algorithm was implemented using Python scripts. To illustrate the performance of the algorithm, in Figure 11, the forecasted profiles for all 500 EVs for two days (weekdays and weekends) across the two seasons were compared with the measured profiles. These values were scaled to the maximum values of the measured data. The curves generally followed the same trend, reaching their peak at approximately the same time. By dividing the curve into two periods, before and after the peak, if the model overestimates the total active power of EVs charging on one side, it underestimates the other.

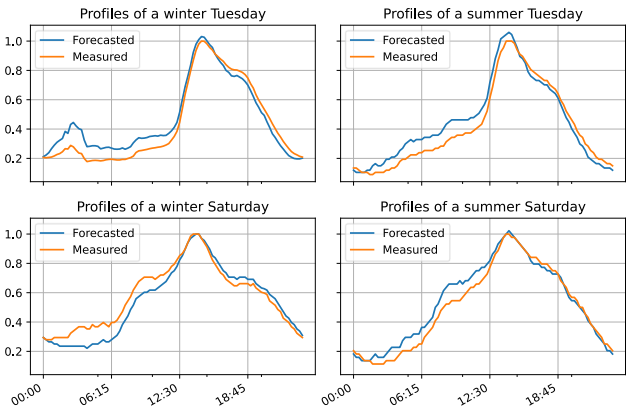

Figure 11. Model performance – Comparison between forecasted and measured profiles.

*B. Application case studies: Results and discussions*

The resulting profiles were integrated into the DN planning tool Cymdist used by HQ's planners. The selected network shown in Figure 12 is mainly composed of long semi-rural lines and contains 3,500 MV/LV transformers, 16000 customers, and 15 feeders. An analysis of different EV penetration rates in the DN of one HQ's DN is performed. The solid lines represent three-phase feeders, dashed lines represent single-phase branches, and dots represent customers.

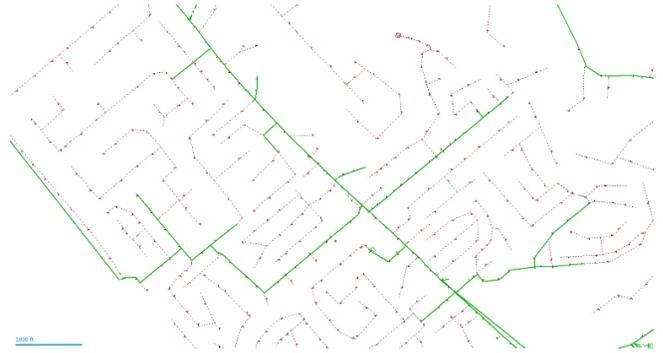

Figure 12. Snapshot of studied network.

Full bottom-up modeling was performed with uncontrolled charging for 30, 50, and 70% EV penetration scenarios. Load demand is assigned based on AMI data available on winter Wednesdays for 100% of customers. The resulting active power at the main breaker of feeder 15 for Phase A is shown in Figure 13. The results were scaled to the maximum value of the base case with no EVs penetration. Without EVs, the maximum consumption happened at 7:00. After introducing EVs, a local peak was always observed in the morning. However, the ultimate peak shifted towards the afternoon and increased with the penetration rate.

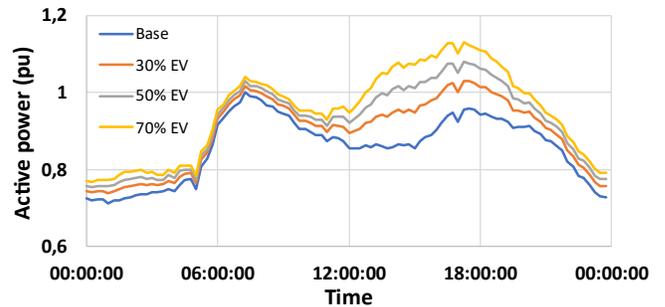

Figure 13. Active power for different EV penetration levels [Feeder 15 phase A]

Distribution-network planners typically use a single-peak network during their planning process. The provided profiles specify that the peak time moved towards the end of the afternoon. In addition, they can use multiple temporally dependent and consecutive networks to perform a quasi-static analysis of the power system [21] – [22]. The simulation results demonstrated that no overloading violations were reached for any scenario. However, the number of undervoltage violations drastically increased with the penetration rate. As shown in Figure 14, the lines and transformers were the most impacted elements of the DS. For the fuses, the effect is negligible. The undervoltage increased slightly with the penetration of EVs. This impact can be damaging because there is only one breaker per feeder. Network enhancement is necessary to accommodate these penetrations. Furthermore, granular analyses based on the consumption of each EVs user can also be performed.

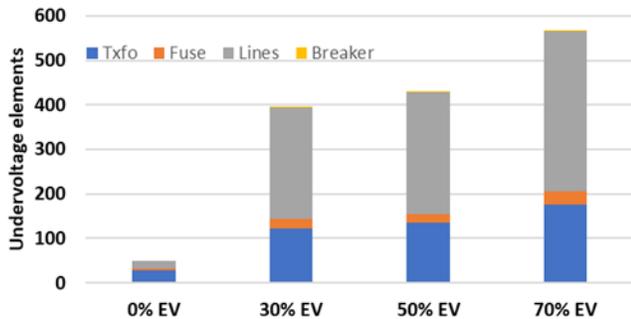

Figure 14. Distribution of voltage violations by electrical component

## V. CONCLUSIONS

In this study, a stochastic approach considering multiple daily charging events, different EV types, and local charging behavior is proposed to better represent individual residential EV charging profiles. The integrated and forecasted model relies on discrete probability functions extracted from residential consumption measurements to capture the realistic local behaviors of EVs users. The EVs charging capacity is inversely related to the number of recharges per day and directly related to the charging duration. Moreover, the number of EVs used to create the charging profiles can have a considerable impact on the charging patterns and parameterization of the model. A case study on the application of the algorithm was also presented. The profiles thus provided can improve the distribution network planning process by performing quasi-static time-series simulations. Distribution-network planners can consider multiple scenarios to find non-wire alternatives and optimal solutions.